\documentclass[12pt]{article}
\usepackage[hypertex]{hyperref}
\usepackage[dvipdfmx]{graphicx} 
\usepackage{graphicx,amsmath,amssymb,amsfonts,cite,bm}

  \def\g{\gamma} \def\G{\Gamma} \def\d{\delta}  \def\e{\epsilon}        \def\m{\mu} \def\n{\nu}             

\def\dg{\dagger} \def\del{\partial} 

\newcommand{\sla}[1]{#1 \!\!\!\! /}

\newcommand{\Lg}{\mathcal{L}}

\newcommand{\row}[2]{ \begin{pmatrix}  #1 & #2   \end{pmatrix}  }
\newcommand{\column}[2]{ \begin{pmatrix}  #1 \\ #2 \\  \end{pmatrix} }
\newcommand{\diag}[2]{ \begin{pmatrix}  #1 & 0 \\ 0 & #2 \\   \end{pmatrix}  }
\newcommand{\offdiag}[2]{ \begin{pmatrix} 0 & #1 \\ #2 & 0 \\   \end{pmatrix} }

\begin{document}

%\begin{titlepage}

\begin{flushright}
STUPP-17-231 
\end{flushright}

\vskip 1.35cm

\begin{center}
{\Large \bf Flavor structure from misalignment of inner products  in noncommutative geometry}

\vskip 1.2cm

Masaki J. S. Yang

\vskip 0.4cm

{\it Department of Physics, Saitama University, \\
Shimo-okubo, Sakura-ku, Saitama, 338-8570, Japan\\
}
%\date{\today}

%%%%%%%%%%%%%%%%%%%%%%%%%%%%%%%%%%%%%%%%%%%%%%%%
\begin{abstract} %%%%%%%%%%%%%%%%%%%%%%%%%%%%%%%%%%%%%%%
%%%%%%%%%%%%%%%%%%%%%%%%%%%%%%%%%%%%%%%%%%%%%%%%

In this letter, we consider an idea 
that induces flavor structure from inner products 
in noncommutative geometry. 
Assuming proper components of vectors $v_{(L,R) i}$ in 
enlarged representation space for fermions, 
we can induce the waterfall texture for Yukawa matrices  
retaining gauge interactions are universal. 
The hierarchy of the Yukawa interactions is a consequence of ``misalignment'' between the vectors $v_{Li}$ and $v_{Rj}$.

%%%%%%%%%%%%%%%%%%%%%%%%%%%%%%%%%%%%%%%%%%%%%%%
\end{abstract} %%%%%%%%%%%%%%%%%%%%%%%%%%%%%%%%%%%%%%%%
%%%%%%%%%%%%%%%%%%%%%%%%%%%%%%%%%%%%%%%%%%%%%%%

\end{center}
%\end{titlepage}

%%%%%%%%%%%%
%\section{Introduction}
%%%%%%%%%%%%

Although the Higgs boson was found at the LHC \cite{Aad:2012tfa,Chatrchyan:2012ufa}, the existence of the particle 
sheds further conundrums, {\it e.g.,} its theoretical origin, the hierarchy problem, and the flavor puzzle.
Among various theories that aim to clarify the origin of the Higgs boson, 
the Yang--Mills--Higgs model in noncommutative geometry (NCG) \cite{Connes:1990qp} is an elegant possibility. 
In this model, the Higgs boson is identified as the gauge boson of the fifth dimension 
which has the noncommutative differential algebra.
In this context of NCG, nontrivial flavor structures are usually introduced {\it by hand} to the distance of the extra dimension, $M\to M \otimes  (Y_{u}, Y_{d}, Y_{e})_{ij} $ in proper representation spaces \cite{Chamseddine:1992kv, Chamseddine:1996zu}. A lot of paper treats the intricate flavor structures in the Standard Model \cite{Sato:1997hv} as one of the ``principles'' or ``axioms''. 
Meanwhile, in the phenomenological region, innumerable theories and models has been proposed to explain the flavor structures. For example, continuous or discrete, hundreds of flavor symmetries \cite{Froggatt:1978nt, Harari:1978yi, Ishimori:2010au}, the flavor textures \cite{Fritzsch:1977vd}, an empirical mass relation \cite{Koide:1983qe}, and so on.

The flavor structures are roughly classified in two types, 
the cascade texture and the waterfall texture in Table 1 \cite{Haba:2008dp}. 
\begin{table} [ht]
\centering
\begin{math}
\begin{array}{|c|c|}
\hline
\begin{pmatrix}
\e & \e & \e \\
\e & \d & \d \\
\e & \d & 1 \\
\end{pmatrix}
&
\begin{pmatrix}
\e^{2} & \e \d & \e  \\
\e \d  & \d^{2} & \d \\
\e & \d & 1 \\
\end{pmatrix}
\\[7pt]
\hline
\text{Cascade} & \text{Waterfall} \\ \hline
\end{array}
\end{math}
\vspace{0.5cm}
\caption{The cascade and waterfall texture, with $1 \gg \d \gg \e$ \cite{Haba:2008dp}. }
\label{obs}
\end{table}
In grand unified models with type-I seesaw mechanism \cite{seesaw}, 
the waterfall one is more phenomenologically desirable \cite{Stech:1999te, Yang:2016crz}. 
It is because the majorana mass matrix of the right-handed neutrinos 
$M_{R} \sim v^{2} Y_{\n}^{T} m_{\n}^{-1} Y_{\n}$ basically shows the waterfall texture 
whichever texture $Y_{\n}$ has. 

Then, in this letter, we consider an idea 
that induces flavor structure from inner products 
in noncommutative geometry. 
Assuming proper components of vectors $v_{(L,R) i}$ in 
enlarged representation space for fermions, 
we can induce the waterfall texture for Yukawa matrices  
retaining gauge interactions are universal. 
The hierarchy of the Yukawa interactions is a consequence of ``misalignment'' between the vectors $v_{Li}$ and $v_{Rj}$. 

The interpretation and origin of the enlarged space are not clear. 
In a toy model with two flavor, we used four times larger one, eight-dimensional inner space. 
%Probably it is the minimal space to induce the waterfall texture. 
%
This idea is similar to Yukawa interactions from wave function overlap 
in theories with extra dimensions \cite{ArkaniHamed:1999dc}. 
Then, perhaps the vectors $v_{(L,R) i}$ can be interpreted as
wave function property of discrete extra dimension 
by solving some equation of motions.

%This paper is organized as follows. 
%In the next section, we review the extended fermionic Lagrangian of the generalized gauge theory in NCG.
%In Sect. 3, the mechanism induces nontrivial flavor structure is presented.
%Section 4 is devoted to conclusions.

%%%%%%%%%%%%%%%%%%%%%%%%%%%%%%%%
%\section{Extended connection and fermionic Lagrangian}
%%%%%%%%%%%%%%%%%%%%%%%%%%%%%%%%

\vspace{18pt}

At the beginning, we briefly review the Higgs mechanism in NCG. 
The following discussions are only presented for the fermionic sector.  
Those of the bosonic sector are found in reviews \cite{Martin:1996wh, Castellani:2000xt}.
The spacetime is considered as $M^{4} \times Z_{2}$,  
 the product of the usual Minkowski space and the two discrete points.
The coordinates are represented by $x^{M} = (x^{\m}, y = \pm)$. 
Operating the exterior derivative $d$ to the relation $y^{2} = 1$, 
an anti-commutative algebra $y \, dy = - dy \, y$ is obtained.
It generates nonzero  Higgs potential. 
%The extension to the $M^{4} \times Z_{N}$ is straightforward.

The exterior derivative of a matrix-formed function $f(x)$ is defined as \cite{Morita:1993dn}:
\begin{align}
 \bm d f \equiv df + d_{5} f \equiv \del_{\m} f dx^{\m} + [D, f] dy. 
\end{align}
Here,
\begin{align}
D = 
\begin{pmatrix}
0 & M \\ 
 M^{\dg} & 0 \\
\end{pmatrix} , 
\end{align}
is the {\it distance matrix} which determines vacuum expectation value (vev) 
and the mass of the Higgs boson. 
Since $M$ is arbitrary parameters, the model still works 
when $M$ is the zero matrix $M = 0$.
This condition leads to the Higgs boson without vev and mass \cite{Yang:2015zoa}. 
Hereafter, we impose $M = 0$ and $\bm d = d$. The nilpotency of $\bm d$ is evident.

The extended connection and chiral fermions are introduced as \cite{Morita:1993dn}:
\begin{align}
\bm A (x) &= 
\begin{pmatrix}
A_{L \m} (x) dx^{\m} &  H (x) dy \\
H^{\dg} (x) dy & A_{R \m} (x) dx^{\m} \\
\end{pmatrix} , 
~~~
\Psi = \column{\psi(x,+)}{\psi(x,-)}
 \equiv \column{\psi_{L}}{\psi_{R}} .
\end{align}

In order to build the fermionic Lagrangian, we define
the Dirac operator for fermions by replacing $(dx^{\m} , dy )$ to $\G^{M} = (\g^{\m}, i \g^{5}) $ 
in $\bm D = d  + \bm A $
\begin{align}
\G^{M} \bm D_{M} %&\equiv [(dx^{\m} , dy ) \to  (\g^{\m}, i \g^{5}) ~{\rm in} ~ \bm D = d + \bm A ] \\ 
& \equiv 
 \g^{\m} \diag{\del_{\m} + A_{L\m}}{\del_{\m} + A_{R\m}} + 
i \g^{5} \offdiag{H}{H^{\dg}} ,
\end{align}
where $\G^{M}$ satisfies the Clifford algebra $\{ \G^{M}, \G^{N} \} = 2 g^{MN}$. 
Rescaling the connections $A_{L,R} \to - i g A_{L,R}$ and $H \to - i g H$, finally
the fermionic Lagrangian is given by
\begin{align}
\Lg_{F} = \bar \Psi i \G^{M} \bm D_{M} \Psi 
&=  
\row{\bar \psi_{L}}{\bar \psi_{R}} 
\begin{pmatrix}
i \sla D_{L} & + i g \g^{5} H \\
- i g \g^{5} H^{\dg} & i \sla D_{R}
\end{pmatrix} 
\column{\psi_{L}}{\psi_{R}} , \label{8}
\\ & = 
\row{\bar \psi_{L}}{\bar \psi_{R}} 
\left[
\begin{pmatrix}
i \g^{\m} \del_{\m} & 0 \\
0 & i \g^{\m} \del_{\m}
\end{pmatrix}
+
g
\begin{pmatrix}
 \g^{\m} A_{L\m} & H \\
H^{\dg} & \g^{\m} A_{R\m}
\end{pmatrix}
\right]
\column{\psi_{L}}{\psi_{R}} . \label{fermionL}
\end{align}
Here, the covariant derivative are $\sla D_{(L,R)} = \g^{\m} (\del_{\m} - igA_{(L,R) \m})$.   
In the last line the $i \g^{5}$ is removed by a proper chiral transformation. 
The overall size of Yukawa interactions are rescaled by the normalization of $H$, to obtain the canonical kinetic term $(D_{\m} H )^{\dg} D^{\m} H$ in the extended field strength $F^{MN} F_{MN}$. 
Note that the vector space in Eqs.~(\ref{8}),(\ref{fermionL}) is not the space of the  Dirac matrices but discrete $Z_{2}$ points in $M^{4} \times Z_{2}$.

%%%%%%%%%%%%%%%%%%%%%%%
%\section{Flavor structure from inner products}
%%%%%%%%%%%%%%%%%%%%%%%
\vspace{18pt}

A problem in this model is the flavor structure. 
Since the gauge and Higgs bosons are unified, 
 naively the Yukawa matrices should be universal or the identity matrix $1_{3}$. 
Furthermore, it is also difficult to induce nontrivial flavor structures 
retaining the universal gauge coupling. 

For example, we can introduce a flavor structure 
by the redefinition of the fermion fields
\begin{align}
\column{\psi_{L i}}{\psi_{R i}} ' = 
\column{V_{L ij}\psi_{L j}}{V_{Rij} \psi_{R j}} .
\end{align}
However, Since the universality of gauge interaction requires $V^{\dg}_{(L,R) ik} V_{(L,R) kj} = \d_{ij}$,  
the form of $V_{(L,R) ij}$ is only restricted to the unitary matrices. 
The Yukawa matrix is found to be
\begin{align}
Y_{ij} = V_{L ik}^{\dg} V_{R kj} ,
\end{align}
and no hierarchy is induced. 
%
%Then, it is an issue to induce nontrivial flavor structures retaining the universality. 
In the following, we show a solution of this point by extending the representation spaces of femions.

%%%%%%%%%%%%%%
%\subsection{The basic idea}
%%%%%%%%%%%%%%
\vspace{12pt}

Here, we describe an idea that induces nontrivial flavor structures. 
Each fermion is assumed to have characteristic vectors $v_{L,R}$ 
in enlarged representation space: 
\begin{align}
\column{\Psi_{L}}{\Psi_{R}} = \column{v_{L} \psi_{L}}{v_{R} \psi_{R}} .
\end{align}
The interactions between bosons and fermions in~(\ref{fermionL}) are rewritten by 
inner products of $v_{L,R}$:
\begin{align}
\Lg_{I} & \equiv 
\row{\bar \Psi_{L} }{\bar \Psi_{R}}
g 
\begin{pmatrix}
\g^{\m} A_{L\m} & H \\
H^{\dg} & \g^{\m} A_{R\m} 
\end{pmatrix}
\column{\Psi_{L}}{\Psi_{R}} 
\\ & = 
\row{\bar \psi_{L} }{\bar \psi_{R} }
g
\begin{pmatrix}
\g^{\m} {A_{L\m} (v_{L}^{\dg}, v_{L})} & {H (v_{L}^{\dg}, v_{R})} \\
 {H^{\dg} (v_{R}^{\dg}, v_{L})} &   \g^{\m} {A_{R\m} (v_{R}^{\dg}, v_{R})}
\end{pmatrix}
\column{\psi_{L}}{\psi_{R}}  .
\end{align}
For example, in a four dimension space, vectors $v_{L,R}$ are assumed to be  
\begin{align}
v_{L} & = 
\begin{pmatrix}
{\sqrt{1 - 2 \e^{2}_{L}}} & {0}  & {\e_{L}} & {\e_{L}}
\end{pmatrix}^{T} ,  \\
v_{R} & = 
\begin{pmatrix}
{0}  & {\sqrt {1 - 2 \e^{2}_{R}}} & {\e_{R}} & {\e_{R}}
\end{pmatrix}^{T} ,
\end{align}
with small parameters $1 \gg \e_{L,R}$. 
Using the inner products 
\begin{align}
(v_{L}^{\dg} , v_{L}) = (v_{R}^{\dg} , v_{R}) = 1, ~~~ 
(v_{L}^{\dg} , v_{R}) = 2 \e_{L}^{*} \e_{R} , 
\end{align}
the interactions are found to be
\begin{align}
\Lg_{I} & = 
\row{\bar \psi_{L} }{\bar \psi_{R} }
\begin{pmatrix}
g \g^{\m} A_{L\m} & 2g \e_{L}^{*} \e_{R}  H \\
2g \e_{R}^{*} \e_{L} H^{\dg} &  g \g^{\m}  A_{R\m} 
\end{pmatrix}
\column{\psi_{L}}{\psi_{R}} .
\end{align}
Therefore we obtain $g_{L} = g_{R} = g$ , and a nontrivial Yukawa coupling $ y = 2g \e_{L}^{*} \e_{R}$.
The difference between the gauge couplings $g_{L,R}$ 
can also induced by the overall redefinition of $v_{L,R}$. 

%%%%%%%%%%%%%%%%
%\subsection{Toy model with two flavor}
%%%%%%%%%%%%%%%%
\vspace{12pt}

This idea can be easily extended to a toy model with two flavor. 
The extension to three flavor $N_{f} = 3$ is also straightforward. 
The chiral fermions in $M^{4} \times Z_{2}$ 
will have the vectors 
\begin{align}
\column{\Psi_{L i}}{\Psi_{R i}} = \column{v_{L i} \psi_{L i}}{v_{R i} \psi_{R i}} ,
\end{align}
with flavor indices $i,j = 1,2$.
The interaction Lagrangian is written as
\begin{align}
\Lg_{I} &=  
\row{\bar \Psi_{L i}}{\bar \Psi_{R i}}
g 
\begin{pmatrix}
\g^{\m} A_{L\m} & H \\
H^{\dg} & \g^{\m} A_{R\m} 
\end{pmatrix}
\column{\Psi_{L i}}{\Psi_{R i}}
\\ & = 
\row{\bar \psi_{L i} }{\bar \psi_{R i} }
g
\begin{pmatrix}
\g^{\m} {A_{L\m} (v_{L i}^{\dg}, v_{L j})} & {H (v_{L i}^{\dg}, v_{R j})} \\
 {H^{\dg} (v_{R i}^{\dg}, v_{L j})} &   \g^{\m} {A_{R\m} (v_{R i}^{\dg}, v_{R j})}
\end{pmatrix}
\column{\psi_{L}}{\psi_{R}} .
\end{align}
The vectors in a eight dimension %(probably it is the minimal space to induce the waterfall texture) 
are assumed to be
\begin{align}
v_{L1} & = 
\begin{pmatrix}
c_{\e_{L}} & 0 & 0 & 0 & \e_{L} & \e_{L} & 0 & 0 
\end{pmatrix}^{T} , \\
v_{L2} & = 
\begin{pmatrix}
0 & c_{\d_{L}} & 0 & 0 & 0 & 0 & \d_{L} & \d_{L}  
\end{pmatrix}^{T} , \\
v_{R1} & =  
\begin{pmatrix}
0 & 0 &  c_{\e_{R}} & 0 & \e_{R} & 0 & \e_{R} & 0 
\end{pmatrix}^{T} , \\
v_{R2} & =  
\begin{pmatrix}
0 & 0 & 0 & c_{\d_{R}} & 0 & \d_{R} & 0 & \d_{R}  \\
\end{pmatrix}^{T} .
\end{align}
Here, $c_{X} \equiv \sqrt{1 - 2 X^{2}}$ with small parameters $1 \gg \d_{L,R}\gg \e_{L,R}$. 
The inner products are found to be 
\begin{align}
(v_{L i}^{\dg} , v_{L j}) = (v_{R i}^{\dg} , v_{R j}) = \d_{ij}, ~~~ 
(v_{L i}^{\dg} , v_{R j}) = 
\begin{pmatrix}
\e_{L}^{*} \e_{R} & \e_{L}^{*} \d_{R} \\
\d_{L}^{*} \e_{R} & \d_{L}^{*} \d_{R}
\end{pmatrix} \equiv y_{ij} .
\end{align}
Therefore, the interaction Lagrangian is rewritten as 
\begin{align}
\Lg_{I} = 
\row{\bar \psi_{L i} }{\bar \psi_{R i} }
g
\begin{pmatrix}
\d_{ij} \g^{\m} A_{L\m} & y_{ij} H \\
y_{ij}^{\dg} H^{\dg} & \d_{ij} \g^{\m} A_{R\m}
\end{pmatrix}
\column{\psi_{L}}{\psi_{R}} ,
\end{align}
and the waterfall texture in Table 1 is induced. 
The hierarchy of the Yukawa interactions is a consequence of ``misalignment'' between the vectors $v_{Li}$ and $v_{Rj}$. 
Moreover, if we adjust the parameters in $v_{(L,R) i}$, 
the determinant of $y_{ij}$ will be finite. For example, a redefined $v_{R1}$
\begin{align}
v_{R1}' =
\begin{pmatrix}
0 & 0 & 2\e_{R} & 0 & \e_{R} & 0 &  d_{\e_{R}} & 0
\end{pmatrix}^{T} , ~~~ 5 \e_{R}^{2} + d_{\e_{R}}^{2} = 1 ,
\end{align}
leads to 
\begin{align}
(v_{L i} , v_{R j}) = 
\begin{pmatrix}
2 \e_{L}^{*} \e_{R} & \e_{L}^{*} \d_{R} \\
\d_{L}^{*} \e_{R} & \d_{L}^{*} \d_{R}
\end{pmatrix} \equiv y'_{ij} .
\end{align}
Then, the Yukawa matrix $y_{ij}'$ has finite determinant and  eigenvalues:
\begin{align}
\det y_{ij}' = \e_{L}^{*} \e_{R} \d_{L}^{*} \d_{R} \neq 0, ~~~~~ y_{1} \simeq \e_{L}^{*} \e_{R}, ~~~ y_{2} \simeq \d_{L}^{*} \d_{R}.
\end{align}
%

%%%%%%%%%%%%%%%%%%%%
%\section{Conclusions and Discussions}
%%%%%%%%%%%%%%%%%%%%
\vspace{18pt}

To conclude, in this letter, we considered an idea 
that induces flavor structure from inner products 
in noncommutative geometry. 
Assuming proper components of vectors $v_{(L,R) i}$ in 
enlarged representation space for fermions, 
we can induce the waterfall texture for Yukawa matrices  
retaining gauge interactions are universal. 
The hierarchy of the Yukawa interactions is a consequence of ``misalignment'' between the vectors $v_{Li}$ and $v_{Rj}$. 

The interpretation and origin of the enlarged space are not clear. 
In a toy model with two flavor, we used four times larger one, eight-dimensional inner space. 
%Probably it is the minimal space to induce the waterfall texture. 
%
This idea is similar to Yukawa interactions from wave function overlap 
in theories with extra dimensions \cite{ArkaniHamed:1999dc}. 
Then, perhaps the vectors $v_{(L,R) i}$ can be interpreted as
wave function property of discrete extra dimension 
by solving some equation of motions. 

{\bf Acknowledgements:} This study is financially supported by the Iwanami Fujukai Foundation, and
Seiwa Memorial Foundation. 

%\bibliographystyle{bib/h-physrev50}
%\bibliography{bib/NCG,bib/democratic,bib/flavor}

\end{document}